\documentclass[prd,twocolumn,superscriptaddress,floatfix,amsmath,amssymb,amsfonts,nofootinbib,longbibliography]{revtex4-2}

\usepackage{float} 
\usepackage{scalerel}
\usepackage[normalem]{ulem}
\usepackage[english]{babel}
\usepackage{graphicx}
\usepackage{dcolumn}
\usepackage{bm}
\usepackage{blindtext}
\usepackage{verbatim}
\usepackage{relsize}
\usepackage{mathrsfs}
\usepackage{musicography}
\usepackage{amsmath}
\usepackage{blindtext}
\usepackage{cancel}
\usepackage{physics}
\usepackage{epstopdf}
\usepackage{mathtools}
\usepackage{blindtext}
\usepackage{tensor}
\usepackage{color}
\usepackage[usenames,dvipsnames]{pstricks}
\usepackage{epsfig}
\usepackage{pst-grad} 
\usepackage{pst-plot} 
\usepackage{hyperref}
\usepackage{verbatim}
\usepackage{slashed}
\usepackage{dsfont}
\usepackage{upgreek}

\newcommand{\mf}{\mathsf}

\newcommand{\ii}{\mathrm{i}}

\newcommand{\tc}[1]{\textsc{#1}}

\allowdisplaybreaks[1] 

\begin{document}

\title{The stress-energy tensor of an Unruh-DeWitt detector}

\author{T. Rick Perche}
\email{trickperche@perimeterinstitute.ca}

\affiliation{Perimeter Institute for Theoretical Physics, Waterloo, Ontario, N2L 2Y5, Canada}
\affiliation{Department of Applied Mathematics, University of Waterloo, Waterloo, Ontario, N2L 3G1, Canada}
\affiliation{Institute for Quantum Computing, University of Waterloo, Waterloo, Ontario, N2L 3G1, Canada}

\author{J. P. M. Pitelli}
\email{pitelli@unicamp.br}
\affiliation{Departamento de Matem\'atica Aplicada, Universidade Estadual de Campinas,
13083-859 Campinas, S\~ao Paulo, Brazil}

\author{Daniel A.\ T.\ Vanzella}
\email{vanzella@ifsc.usp.br}
\affiliation{Instituto de F\'\i sica de S\~ao Carlos,
Universidade de S\~ao Paulo, Caixa Postal 369, CEP 13560-970, 
S\~ao Carlos, SP,  Brazil}

\begin{abstract}
We propose a model for a finite-size particle detector, which allows us to derive its stress-energy tensor. This tensor is obtained from a covariant Lagrangian that describes not only the quantum field that models the detector, $\phi_{\tc{d}}$, but also the systems responsible for its localization: a complex scalar field $\psi_{\tc{c}}$, and a perfect fluid. The local interaction between the detector and the complex field ensures the square integrability of the detector modes, while the fluid serves to define the spatial profile of $\psi_{\tc{c}}$, localizing it in space. We then demonstrate that, under very general conditions, the resulting energy tensor---incorporating all components of the system---is physically reasonable and satisfies the energy conditions.
\end{abstract}
\maketitle

\section{Introduction}

In the formalism of quantum theory, measuring apparatuses  play a special role.  Serving as windows  through which we can have  glimpses of Nature at its most fundamental  level, they  lie at the interface between the quantum and the classical worlds. As such, going beyond their highly idealized descriptions and modeling their dynamics and behavior may help to investigate questions related to the quantum-classical frontier. In particular, when it comes to  questions involving gravity and its quantum aspects,  a model of the apparatus from which one can derive a sensible stress-energy tensor is essential, since this is the observable through which the background spacetime perceives the apparatus and the measuring process. In this direction, we propose a model for a finite-size particle detector which allows us to derive its stress-energy tensor.

Since the dawn of classical physics, the concepts of {\it energy}  and {\it momentum} have played a central role in our understanding of physical processes and fundamental principles. These concepts were eventually understood to be intimately related to (symmetries w.r.t.)~the more basic notions of {\it time} and {\it space}, respectively. Special Relativity brought a paradigm shift where the
relative notions of space and time were interwoven into an absolute four dimensional  ``arena''---the {\it  spacetime}, on which  physical processes unfold according to {\it local} laws. Energy and momentum  were then combined into  the {\it stress-energy-momentum tensor} (or simply stress-energy tensor, or energy-momentum tensor),  a rank-two tensor which encodes how energy and momentum  flow  in the system and how they are exchanged with ``external  agents.'' Then, a second paradigm shift, brought by General Relativity, took the connection between energy/momentum and time/space to a new level by attributing to the stress-energy tensor the fundamental role of sourcing deformations on the fabric of spacetime (i.e., {\it gravity}), elegantly making {\it local} energy-momentum conservation a direct consequence of the spacetime dynamics and general covariance. For this reason, the stress-energy tensor  is the
most important  object of a system when it comes to its interplay with gravity.

Although the picture above is complete at the classical level,  the question of  how  the spacetime is affected by a {\it quantum} system---more  precisely, a system in a quantum state with no well-defined energy-momentum distribution---lies beyond our current (theoretical and experimental) knowledge. In face of this, it is conceivable that following closely the ``energetics'' of the measuring process by which
the quantum realm connects to the classical world---which is completely hidden in the highly idealized description of measurements in the usual  formalism---may lead to interesting insights. And in order to accomplish this, a localized measuring apparatus must be modeled in a way that a sensible stress-energy tensor can be ascribed to it. 

One way of implementing quantum measurements in relativistic settings was put forward by Fewster and Verch~\cite{FewsterVerch,fewster2} and consists of a measurement theory entirely described in the context of algebraic QFT. Although this framework is fully covariant and mathematically consistent, it is not entirely clear how one applies it to describe a localized system. On the other hand, an operational approach to measurements in QFT was proposed in~\cite{chicken}, describing the measurement apparatus as an Unruh-DeWitt (UDW) detector~\cite{Unruh1976,DeWitt} that locally couples to a quantum field. This particle detector approach to measurements has the advantages of naturally describing localized systems and being in direct analogy with physical setups~\cite{eduardoOld,Pozas2016,Nicho1,richard,neutrinos,pitelligraviton}. However, these detector models are not fully compatible with relativity, displaying breakdown of causality~\cite{martin-martinez2015,PipoFTL,mariaPipoNew} and/or covariance~\cite{us,us2} on length scales below the size of the detector.

One way of connecting the Fewster-Verch framework with the particle detector approach is to notice that a UDW detector admits a quantum field theoretic description~\cite{Unruh1976,QFTPD}, for instance as a discrete-energy mode of a quantum field $\hat{\phi}_\tc{d}$ confined by a
{\it prescribed} classical  potential $V({\mf x})$. Although this gives rise to a field-theoretic description of a localized measurement apparatus, a non-dynamical potential gives rise to a stress-energy tensor incompatible with general covariance, implying that models of this type are also unsuited to investigate some matters of principle.

Here, we present, a {\it relativistically-consistent}  particle detector, modeling all its ``constituents'' and deriving the corresponding stress-energy tensor. We build on the description of a UDW detector as a discrete-energy mode of a quantum field $\hat{\phi}_\tc{d}$ under the influence of a confining potential. However, in our framework, the confining potential is not prescribed as a function of the spacetime points but rather as due to interaction with a dynamical self-interacting (complex) field $\psi_\tc{c}$. The spatial profile of $\psi_\tc{c}$ is then determined by its coupling with a perfect fluid. Although seemingly convoluted, this semiclassical description bears some analogy with, e.g., a second-quantized electron (here modeled by $\hat{\phi}_\tc{d}$) bounded in an atom by the classical electromagnetic field (whose role is played by $\psi_\tc{c}$) generated
by the central nucleus (analogous to the perfect fluid).

The paper is organized as follows. In Sec.~\ref{sec:RPD}, we review the standard UDW  model (Subsec.~\ref{sub:locQFTs}) and how it functions as a particle detector (Subsec.~\ref{sub:probing}). In  Sec.~\ref{sec:tmunu}, we review how the stress-energy tensor is obtained for a field theory subject to an external  potential. We discuss the necessity of including the source of the potential in the description. Then,
in Sec.~\ref{sec:model}, we present our model, including in the Lagrangian of the theory the source of the confining potential for the detector field $\hat{\phi}_\tc{d}$ (i.e., the self-interacting field $\psi_{\tc{c}}$) and the perfect fluid which can be seen as sourcing/confining the field $\psi_\tc{c}$. In Sec.~\ref{sec:example}, we present a concrete detector model by choosing a particular example of  self-interacting  potential for the field $\psi_\tc{c}$ and equation of state for the perfect-fluid. Finally, in Sec.~\ref{sec:tmunuexample}, we obtain the stress-energy tensor of this concrete model. Sec.~\ref{sec:conclusion} is dedicated to final
comments and conclusions. Throughout the text we use metric signature $(-,+,+,+)$ and natural units in which $\hbar = c=1$.

\section{Relativistic Particle Detectors}
\label{sec:RPD}

In this section we briefly review the description first conceived in Ref.~\cite{Unruh1976} and later explored in Ref.~\cite{QFTPD}, where a localized  quantum field $\hat{\phi}_\tc{d}$ is used to probe a free quantum field $\hat{\phi}$. The localization of the field $\hat{\phi}_\tc{d}$ is specified by the addition of a classical confining potential $V(\mf x)$. This potential is responsible for the creation of  localized modes for the field $\hat{\phi}_\tc{d}$. These modes interact with the field $\hat{\phi}$, giving rise to a relativistic localized probe.

\subsection{Localized quantum fields}
\label{sub:locQFTs}

The Lagrangian for a scalar field $\phi_\tc{d}$ subjected to an external potential $V(\mf x)$ in 3+1 Minkowski spacetime is given by
\begin{equation}
    \mathcal{L} = -\frac{1}{2} \partial_\mu \phi_\tc{d}\partial^\mu \phi_\tc{d} - \frac{m_\tc{d}^2}{2}\phi_\tc{d}^2 - \frac{V(\mf x)}{2}\phi_\tc{d}^2.
    \label{lagrangian phid}
\end{equation}
The mode solutions of the field $\phi_\tc{d}$ will be localized provided that the potential $V(\mf x)$ is confining. For simplicity, we will restrict ourselves to the case where the potential $V(\mf x)$ is time independent in a given inertial reference frame, so that one can find a basis of solutions for the field $\phi_\tc{d}$ composed  by eigenfunctions of the form $e^{\pm i \omega t}\Phi(\bm x)$. We then assume that $V(\mf x) = V(\bm x)$ in an inertial frame with coordinates $\mf x=(t,\bm x)$. Furthermore, in order to find normalized modes for the field, we also assume that the potential $V(\bm x)$ is such that the spectrum of the operator
\begin{equation}
    L = -\nabla^2 + V(\bm x)
\end{equation}
contains a discrete set {of} eigenvalues $\mu_{\bm n}$, ${\bm n}\in \mathcal{A}\subset \mathbb{N}$, bounded from below, in addition to a continuous spectrum\footnote{This is exactly what happens in the case of the Hydrogen atom, where the operator $- \nabla^2 + \frac{e^2}{r}$ has infinitely many discrete negative eigenvalues $-13.6\textrm{eV}=E_0<E_1<E_2<\cdots<0$, and a continuous spectrum $E_{\bm k}\geq 0$ that can be parametrized by $\bm k\in \mathbb{R}^3$.} parametrized by $\nu_{\bm k}$, with $\bm{k}\in\mathbb{R}^3$. 

Let
\begin{equation}
    L[\Phi_{\bm n}(\bm x)] = \mu_{\bm n}\Phi_{\bm n}(\bm x),
\end{equation}
with $\Phi_{\bm n}(\bm x) \in L^2(\mathbb{R}^3)$, and let $v_{\bm k}(\bm x)$ be the ``generalized'' eigenfunctions in the continuous spectrum of $L$, so that
\begin{equation}
    L[v_{\bm k}(\bm x)] = \nu_{\bm k}v_{\bm k}(\bm x).
\end{equation}

The quantum field can then be written as
\begin{equation}
\begin{aligned}
    \hat{\phi}_\tc{d}(\mf x) =& \sum_{\bm n} \left(e^{- \ii \omega_{\bm n} t}\Phi_{\bm n}(\bm x) \hat{a}_{\bm n} + e^{\ii \omega_{\bm n} t}\Phi^*_{\bm n}(\bm x) \hat{a}^\dagger_{\bm n}\right)\\
    &+ \int \dd^3 \bm k \left(e^{- \ii w_{\bm k} t}v_{\bm k}(\bm x) \hat{b}_{\bm k} +e^{\ii w_{\bm k} t}v^*_{\bm k}(\bm x) \hat{b}^\dagger_{\bm k} \right){,}
\end{aligned}
 \label{decomposition}
\end{equation}
where
\begin{align}
    \omega_{\bm n} &= \sqrt{m_{\tc{d}}^2 + \mu_{\bm n}}, &&& w_{\bm k} &= \sqrt{m_{\tc{d}}^2 + \nu_{\bm k}},
\end{align}
$\Phi_n(\bm{x})$ is square-integrable and localized around the minima of the potential $V(\bm x)$.  We assume that
the potential $V({\bm x})$ is such that $\mu_{{\bm n}},\nu_{\bm k} > -m_\tc{d}^2$ for all ${{\bm n}}\in{\cal A}$ and ${\bm k}\in {\mathbb R}^3$, so that all these eigenfunctions are stationary solutions\footnote{ If $V({\bm x})$ is sufficiently negative on a sufficiently large region, there will exist eigenfunctions with $\mu_n < -m_\tc{d}^2$. These are unstable solutions of the form $e^{\pm |\omega_n| t}\Phi_n(\bm x)$. These unstable modes behave like eigenstates of the so called inverted harmonic oscillator (IHO), whose Hamiltonian has the  form $H_{\textrm{IHO}}=\frac{p^2}{2m}-\frac{\omega^2 x^2}{2}$. This Hamiltonian corresponds to the usual harmonic oscillator with the correspondence $\omega\to i\omega$. This gives rise to generalized eigenstates with complex eigenfrequencies.  
An interesting scenario where such an instability appears is that of a nonminimally-coupled scalar field in the
background spacetime of a compact astrophysical object---which leads to the phenomena of {\it spontaneous scalarization} in the classical regime~\cite{DEF1993,PCBRS2011} (see also Ref.~\cite{RevSpSc2024} and references therein) 
and of {\it vacuum awakening} in the quantum regime~\cite{LMV2010,LLMV2012}.
For quantization of fields with such unstable solutions, see, e.g., Refs.~\cite{SSW1940,Fulling1976,LV2010,Lima2013,Lima2016,pitelli}.}.
Upon proper normalization of the modes $\Phi_n({\bm x})$ and $v_{\bm k}({\bm x})$ (which sets the scale of quantum fluctuations),
the creation and annihilation operators $\hat{a}_{\bm n}$, $\hat{a}_{\bm n}^\dagger$, $\hat{b}_{\bm k}$, $\hat{b}_{\bm k}^\dagger$ then satisfy the canonical commutation relations
\begin{align}
    [\hat{a}_{\bm n}, \hat{a}_{\bm n'}^\dagger] &= \delta_{\bm n \bm n'}\\
    [\hat{b}_{\bm k}, \hat{b}_{\bm k'}^\dagger] &= \delta^{(3)}(\bm k - \bm k')
\end{align}
and define a vacuum state $\ket{0}$ through the condition $\hat{a}_{\bm n} \ket{0} = \hat{b}_{\bm k}\ket{0} = 0$. Notice the discrete versus continuous canonical commutation relations, according to whether the associated modes are in the discrete or continuous spectrum of the operator $L$. 

The Fock space $\mathcal{F}$ for the field in the representation associated with $\ket{0}$ is given by
\begin{equation}
    \mathcal{F} = \left(\bigotimes_{{\bm n}{\in \cal A}} H_{\bm n}\right)\otimes \mathcal{F}_\text{cont},
\end{equation}
where the spaces $H_{\bm n}$ and $\mathcal{F}_\text{cont}$ can be explicitly written as
\begin{equation}
\begin{array}{rcl}
    H_{\bm n}&  =  &  \text{span}\left(\{(\hat{a}^\dagger_{\bm n})^m \ket{0_{\bm n}}:m\in \mathbb{N}\}\right),\\
     \mathcal{F}_\text{cont}&  = &  \text{span}\left(\{\hat{b}^\dagger_{{\bm k}_1}\dots \hat{b}^\dagger_{{\bm k}_m} \ket{0_\text{cont}}:m\in \mathbb{N},\,\,\bm{k}_j\in \mathbb{R}^3\}\right).
\end{array}
\end{equation}
The decomposition above can be made because the vacuum of the field theory decomposes as 
\begin{equation}
    \ket{0} = \left(\bigotimes_{\bm n} \ket{0_{\bm n}}\right)\otimes \ket{0_{\text{cont}}},
\end{equation}
where the states $\ket{0_{\bm n}}$ correspond to no excitation in the corresponding discrete modes, and the state $\ket{0_\text{cont}}$ is the vacuum associated to the continuous modes.

\subsection{Probing a quantum field with a localized QFT}
\label{sub:probing}

In the context of locally probing a quantum field $\hat{\phi}$, one often considers a localized quantum system that probes  $\hat{\phi}$ in a finite region of spacetime. This allows one to study numerous quantum information protocols in the context of QFT, such as entanglement harvesting~\cite{Valentini1991,Reznik1,reznik2,Salton:2014jaa,Pozas-Kerstjens:2015,HarvestingQueNemLouko,Pozas2016,HarvestingSuperposed,Henderson2019,bandlimitedHarv2020,mutualInfoBH,threeHarvesting2022}, quantum energy teleportation~\cite{teleportation,teleportation2014,nichoTeleport,teleportExperiment}, quantum collect calling~\cite{Jonsson2,collectCalling,PRLHyugens2015}, among others~\cite{KojiCapacity,Ahmadzadegan2021}.

The usual tool employed to locally probe quantum fields are the so-called Unruh-DeWitt particle detector models, introduced in~\cite{Unruh1976,DeWitt}. These models come in many shapes and forms~\cite{Unruh-Wald,eduardoOld,Pozas-Kerstjens:2015,neutrinos,generalPD}, but a concrete realization is to consider the detector to be a quantum harmonic oscillator following a worldline in Minkowski spacetime, with frequency $\Omega$ and creation and annihilation operators $\hat{a}$ and $\hat{a}^\dagger$. The detector then interacts with a scalar quantum field $\hat{\phi}(\mf x)$ according to the interaction Hamiltonian density
\begin{equation}\label{eq:UDWHO}
    \hat{h}_I(\mf x) = \lambda \Lambda(\mf x) (e^{-\ii \Omega t}\hat{a} + e^{\ii \Omega t}\hat{a}^\dagger)\hat{\phi}(\mf x).
\end{equation}
In this context $\lambda$ is a coupling constant and $\Lambda(\mf x)$ is a spacetime smearing function, whose profile define the localization of the interaction in spacetime. This effective model can be derived when one considers the interaction of a quantum system (detector) with an external quantum field $\hat{\phi}$. The spatial localization of the interaction is usually defined by the wavefunctions that define the detector~\cite{Unruh-Wald,generalPD}.

However, the UDW particle detector models are not the only way in which one could conceive locally probing a quantum field. As a matter of fact, when first envisioned by Unruh in~\cite{Unruh1976}, the field was probed using a localized quantum field in a box. In a way, this formulation is more satisfactory, as the internal dynamics of the probe are inherently relativistic. However, in~\cite{QFTPD} it was shown that results obtained by using localized quantum fields or particle detectors are equivalent to leading order in perturbation theory. This allows one to extrapolate any result using UDW detectors to a relativistic setting, and vice versa. In the context of this paper, this analogy will allow us to use  such relativistic description to attempt to formulate the stress-energy tensor of a particle detector.

We will now briefly review how the analogy between UDW detectors and localized quantum fields takes place. The  system that models the interaction of the localized field $\hat{\phi}_\tc{d}$ with an external massless quantum field $\hat{\phi}$ can be prescribed according to the Lagrangian density
\begin{align}
    \mathcal{L} = & -\frac{1}{2} \partial_\mu \phi_\tc{d}\partial^\mu \phi_\tc{d} - \frac{m_\tc{d}^2}{2}\phi_\tc{d}^2 - \frac{V(\mf x)}{2}\phi_\tc{d}^2 \nonumber\\
    &- \frac{1}{2}\partial_\mu \phi \partial^\mu \phi - \lambda\zeta(\mf x)\phi_\tc{d}\phi,
\end{align}
where the last term in the expression above allows for the detector field $\hat{\phi}_\tc{d}$ to couple to the external field $\phi$ in a localized region of spacetime, determined by the profile of the function $\zeta(\mf x)$ with coupling strength controlled by the real parameter $\lambda$. The results of~\cite{QFTPD} then show that, to leading order in $\lambda$, results obtained by considering a non-relativistic particle detector and the field theoretic model above (restricted to a single mode excitation) are equivalent. This result also holds when multiple detectors are considered~\cite{FullHarvesting}.

In essence, each individual discrete mode of the field $\phi_\tc{d}$ acts as a localized particle detector. More specifically, in the setup of Subsection~\ref{sub:locQFTs}, given $\bm n \in \mathcal{A}$, one can compute the time evolution of a state of the form \mbox{$\hat{\rho}_0 = \hat{\rho}_{\tc{d},0}\otimes \hat{\rho}_\phi$}, where $\hat{\rho}_\phi$ is the state of the free field $\hat{\phi}(\mf x)$ and $\hat{\rho}_{\tc{d},0} = \hat{\rho}_{{\bm n}^c}\otimes \hat{\rho}_{\bm n,0}$, with $\hat{\rho}_{{\bm n},0}$ a state in $H_{{\bm n}}$ and $\hat{\rho}_{{\bm n}^c}$ a state in the remaining portion of $\mathcal{F}$, which we denote by $\mathcal{F}_{{\bm n}}^c$. Time evolution of this state is prescribed by the interaction Hamiltonian density
\begin{equation}\label{eq:HIfield}
    \mathcal{H}_I(\mf x) = \lambda \zeta(\mf x) \hat{\phi}_{\tc d}(\mf x) \hat{\phi}(\mf x),
\end{equation}
which produces the interaction time evolution operator $\hat{U}_I$. The results of~\cite{QFTPD} are that, to leading order in $\lambda$, the final state of the mode ${\bm n}$,
\begin{equation}
    \hat{\rho}_{{\bm n}} = \tr_{\phi,\mathcal{F}_{{\bm n}}^c}(\hat{U}_I\hat{\rho}_0\hat{U}_I^\dagger),
\end{equation}
evolves exactly in the same way that it would using the interaction Hamiltonian of Eq.~\eqref{eq:UDWHO} with the replacements $\hat{a}\mapsto \hat{a}_{ {\bm n}}$, $\Omega\mapsto \omega_{{\bm n}}$, and $\Lambda(\mf x)\mapsto \zeta(\mf x) \Phi_{{\bm n}}(\bm x)$, so that it behaves exactly like a harmonic oscillator UDW detector would.

The fact that each localized mode of a quantum field is analogous to a UDW detector is a mere consequence that particle detectors are fundamentally described by relativistic quantum field theories. In particular, the description of a detector as a localized mode of a quantum field is the key to find the stress energy tensor of a UDW detector.

\section{The stress-energy tensor of a localized quantum field}
\label{sec:tmunu}

The classical stress-energy tensor associated with the field $\phi_\tc{d}(\mf x)$,  whose equation of motion is defined by the Lagrangian given by Eq. \eqref{lagrangian phid} is
\begin{equation}\begin{aligned}
    T_{\mu\nu} &= \partial_\mu \phi_{\tc{d}}\partial_\nu \phi_{\tc{d}}\\&-\frac{1}{2}g_{\mu\nu}\Big(\partial_\alpha \phi_{\tc{d}}\partial^\alpha \phi_{\tc{d}} + m_{\tc{d}}^2 \phi_\tc{d}^2 + V(\mf x) \phi_\tc{d}^2\Big).
    \label{TmunuVx}
\end{aligned}\end{equation}
By writing the expression above for the quantum field $\hat{\phi}_\tc{d}$ and taking expected values of the operator-valued stress energy tensor, $\langle \hat{T}_{\mu\nu}\rangle$ (with the appropriate renormalization methods~\cite{birrell_davies,HadamardRenormalization2008}){,} one could claim that this is how the field $\phi_\tc{d}$ gravitates. However, there is an important matter which is not addressed in this description: what is the physical origin of the potential $V(\mf x)$? Whatever is the physical system that sources the potential $V(\mf x)$, both the source and the energy associated to $V(\mf x)$ will contribute non-negligibly to the total stress-energy tensor in the spacetime. 
Moreover, if the source of the localization potential $V({\mf x})$ is not taken into account, the field theory associated to
Eq.~(\ref{TmunuVx}) irredeemably breaks special (and general) covariance. Indeed, the stress-energy tensor of Eq.~\eqref{TmunuVx} is not covariantly conserved:
\begin{equation}
    \partial^\mu T_{\mu\nu} = \partial_\nu \phi_\tc{d}(\partial^\mu \partial_\mu - m_\tc{d}^2 - V(\mf x))\phi_\tc{d} - \frac{1}{2}\phi_\tc{d}^2\partial_\nu V(\mf x),
\end{equation}
which yields $\partial^\mu T_{\mu\nu} = - \frac{1}{2}\phi_\tc{d}^2\partial_\nu V(\mf x)$ on shell. That is, the only potentials that would produce a theory fully compatible with general covariance are constant, being unable to localize the field modes of $\phi_\tc{d}$.

One option would be to consider that $V(\mf x)$ is generated by another scalar field. In that case, $V(\mf x)$ would be replaced by $V(\phi_\tc{c}(\mf x))$ and, for any fixed solution for the field $\phi_\tc{c}$, the potential which would affect $\phi_\tc{d}$ would be a function of $\phi_\tc{c}(\mf x)$. In that case, the full Lagrangian of the theory would be 
\begin{equation}
\begin{aligned}
    \mathcal{L} =& - \frac{1}{2} \partial_\mu\phi_{\tc{d}}\partial^\mu\phi_{\tc{d}} - \frac{m_\tc{d}^2}{2} \phi_{\tc{d}}^2 -\frac{V(\phi_\tc{c})}{2}\phi^2_\tc{d}\\&
    - \frac{1}{2} \partial_\mu\phi_\tc{c}\partial^\mu \phi_\tc{c} - \frac{m_\tc{c}^2}{2}\phi_\tc{c}^2
\end{aligned}
\end{equation}
and the associated stress-energy tensor would be
\begin{equation}
\begin{aligned}
    T_{\mu\nu} =& \partial_\mu \phi_{\tc{d}}\partial_\nu \phi_{\tc{d}}+\partial_\mu \phi_{\tc{c}}\partial_\nu \phi_{\tc{c}}\\&-\frac{1}{2}g_{\mu\nu}\Big[\partial_\alpha \phi_{\tc{d}}\partial^\alpha \phi_{\tc{d}} + m_{\tc{d}}^2 \phi_\tc{d}^2 + V(\phi_\tc{c})\phi_\tc{d}^2\\
    &\:\:\:\:\:\:\:\:\:\:\:\:\:\:\:\:\:\:\:\:\:\:\:\:\:\:\:\:\:\:\:\:+ \partial_\alpha \phi_{\tc{c}}\partial^\alpha \phi_{\tc{c}} + m_{\tc{c}}^2 \phi_\tc{c}^2\Big]{,}
\end{aligned}
\end{equation}
which is covariantly conserved on shell. However, the solutions of the Klein-Gordon equation for $\phi_\tc{c}$ will generally not be confined to a finite region of space and will propagate away. One way to prevent this would be to find a fine tuned solitonic solution for the fields $\phi_\tc{d}$ and $\phi_\tc{c}$. Unfortunately, this would imply that small changes to the system (such as when $\phi_\tc{d}$ interacts with an external field) would likely break the bound system, leading to the fields $\phi_\tc{c}$ and $\phi_\tc{d}$ propagating away.

An alternative way of keeping the field $\phi_\tc{c}$ localized in space would be to consider an external potential $V_\tc{c}(\bm x)$ which localizes it. This would amount to adding a term of the form $- \frac{1}{2}V_\tc{c}(\bm x) \phi_\tc{c}^2$ to the Lagrangian, which would allow for a bound solution for $\phi_\tc{c}$. One would then naturally wonder what is the physical system that sources the potential $V_\tc{c}$ and what are its contributions to the stress-energy tensor. One could, of course, introduce yet another scalar field $\phi_\tc{b}$, and to add another interaction of the form $V(\phi_\tc{b})\phi_\tc{c}^2$, but this would quickly lead us to a rabbit hole, where one would always need to add another localized field to source the effective potential that localizes the previous one.


A possible alternative 
solution to this puzzle (as we will show below) is to introduce matter with two degrees of freedom---a perfect fluid, for instance. One degree of freedom would be responsible for the localization of the field $\phi_{\tc{c}}$ regardless of the state of the field $\phi_\tc{d}$, and the other one would depend on the equations of motion and a boundary condition. As we will see in the next section, the additional matter with two degrees of freedom would allow a stable solitonic solution for the fields $\phi_\tc{c}$ and $\phi_\tc{d}$, maintaining the shape of the solution $\phi_\tc{c}$ (and thus of the localizing potential $V(\phi_\tc{c})$) regardless of the state of $\phi_\tc{d}$.


\section{A consistent semiclassical description for a particle detector}
\label{sec:model}

The description of a localized quantum field which takes into account the physical system that localizes the field can be formulated in terms of a Lorentz invariant action in Minkowski spacetime. The full Lagrangian depending on the scalar field $\phi_\tc{d}$, and a complex\footnote{We chose a complex-valued field so that a time-independent potential and energy-momentum distribution (which basically depends on $|\psi_\tc{c}(\mf x)|^2$)  can be obtained  when $\psi_\tc{c}$ is in a stationary state.} field $\psi_\tc{c}$ that produces an effective potential responsible for the localization of $\phi_{\tc{d}}$ through the potential $V(|\psi_{\tc{c}}|^2)$ and on the fluid configuration  is given by
\begin{equation}
\begin{aligned}
    \mathcal{L} =& - \frac{1}{2} \partial_\mu\phi_{\tc{d}}\partial^\mu\phi_{\tc{d}} - \frac{m_\tc{d}^2}{2} \phi_{\tc{d}}^2 - \frac{\alpha}{2} |\psi_{\tc{c}}|^2\phi_{\tc{d}}^2\\
    &- \partial_\mu\psi_\tc{c}^*\partial^\mu \psi_\tc{c} - m_\tc{c}^2|\psi_\tc{c}|^2 - V_{\tc{c}}(|\psi_\tc{c}|^2)\\&+(1- \mu  |\psi_\tc{c}|^2) \mathcal{L}^{\textrm{fluid}},
    \label{full lagrangian}
\end{aligned}
\end{equation}
where $\mu$ is a coupling constant with units of squared length, $\alpha$ is a dimensionless constant, $m_\tc{d}$ and $m_\tc{c}$ are the masses of the fields $\phi_\tc{d}$ and $\psi_\tc{c}$ and $V_\tc{c}(|\psi_\tc{c}|^2)$ is a self-interaction term for the field $\psi_\tc{c}$.  

The role of the Lagrangian $\mathcal{L}^{\textrm{fluid}}$ in this description is two-fold. It gives rise to the energy-momentum of the fluid and appears explicitly on the equations of motion for $\psi_\tc{c}$ due to the non-minimal coupling between the fluid and the field $\psi_{\tc{c}}$. In this way, the exact form of the on-shell Lagrangian $\mathcal{L}^{\textrm{fluid}}$ turns out to be essential. There are several possible on-shell real Lagrangians (all of them giving rise to the same energy-momentum tensor). The most common are $\mathcal{L}^\textrm{fluid}=P$~\cite{Schutz1970} and $\mathcal{L}^\textrm{fluid}=-\rho$~\cite{Brown1993}, where $P$ and $\rho$ are the proper pressure and the proper energy density of the fluid. The transition between these two Lagrangians is made through the addition of a surface integral in the fluid action, i.e., the Lagrangian is modified by a total derivative term~\cite{Brown1993}. This clearly affects its on-shell value without affecting the equations of motion. Also, by considering the fluid as constituted by particles with fixed rest mass and structure (solitons), the average on-shell Lagrangian turns out to be of the form~\cite{Avelino2018} 
\begin{equation}
    \mathcal{L}^{\textrm{fluid}} = T^{\textrm{fluid}} = -\rho+3P,
\end{equation}
i.e., the trace of the stress-energy tensor 
\begin{equation}
    T^\textrm{fluid}_{\mu\nu}  =  (\rho + P)u_\mu u_\nu+ P g_{\mu\nu}.
\end{equation}
In this case, for the fluid to be modelled by a collection of particles, the equation of state $w=p/\rho$ must satisfy $0\leq w\leq 1/3$. 

The equations of motion for the fields $\phi_\tc{d}$ and $\psi_\tc{c}$ are given  by
\begin{subequations}
      \begin{align}
        & (\Box-m_\tc{d}^2 - \alpha|\psi_\tc{c}|^2)\phi_\tc{d}  = 0,\label{eqmotion1}\\
        &\left(\Box-m_\tc{c}^2- F_\tc{c}(|\psi_\tc{c}|^2) -\mu  \mathcal{L}^{\textrm{fluid}}- \frac{\alpha}{2}\phi_\tc{d}^2\right) \psi_{\tc{c}} = 0,\label{eqmotion2}
      \end{align}
      \label{eqmotion}
    \end{subequations}
where we defined
\begin{equation}
    F_\tc{c}(|\psi_\tc{c}|^2) = \pdv{V_{\tc{c}}}{|\psi_{\tc{c}}|^2}.
\end{equation}
Notice that Eq.~(\ref{eqmotion2}) shows explicitly how the Lagrangian of the fluid $\mathcal{L}^{\textrm{fluid}}$ affects the equations of motion for $\psi_\tc{c}$.

The equations of motion for a perfect fluid minimally coupled to gravity are equivalent to $\nabla^\mu T_{\mu\nu}^{\textrm{fluid}}=0$ along with the conservation of particle number~\cite{Schutz1970}. However, in our case, the fluid is also coupled to matter fields. Hence the equations of motion for the fluid turns out to be given by the conservation of the full stress-energy tensor in the spacetime, i.e.,  $\nabla_\mu T^{\mu\nu}=0$, where 
\begin{widetext}
\begin{equation}
\begin{aligned}\label{eq:fullTmunu}
    T_{\mu\nu} \equiv& - \frac{2}{\sqrt{-g}} \frac{\delta (\sqrt{-g}\mathcal{L})}{\delta g^{\mu\nu}}\\
    =& \partial_\mu \phi_{\tc{d}}\partial_\nu \phi_{\tc{d}}+2\Re(\partial_\mu \psi_{\tc{c}}^*\partial_\nu \psi_{\tc{c}}) -\frac{1}{2}g_{\mu\nu}\Big(\partial_\alpha \phi_{\tc{d}}\partial^\alpha \phi_{\tc{d}} + m_{\tc{d}}^2 \phi_\tc{d}^2 + \alpha |\psi_\tc{c}|^2 \phi_\tc{d}^2\\
    &+ 2\partial_\alpha \psi_{\tc{c}}^*\partial^\alpha \psi_{\tc{c}} + 2m_{\tc{c}}^2 |\psi_\tc{c}|^2 + 2 V_\tc{c}(|\psi_\tc{c}|^2)\Big)+(1-\mu |\psi_\tc{c}|^2)T^\textrm{fluid}_{\mu\nu}.
\end{aligned}
\end{equation}
We can then compute its divergence,

\begin{equation}
\begin{aligned}
    \partial^\mu T_{\mu\nu} =&  \left[\Big(\Box-m_\tc{d}^2 - \alpha|\psi_\tc{c}|^2 \Big)\phi_{\tc{d}}\right]\partial_\nu \phi_{\tc{d}} +2\Re\left\{\left[\Big(\Box-m_\tc{c}^2  -F_\tc{c}(|\psi_\tc{c}|^2)- \mu  \mathcal{L}^\textrm{fluid}- \tfrac{\alpha}{2}\phi_\tc{d}^2\Big)\psi_{\tc{c}}\right]\partial_\nu \psi_{\tc{c}}^*\right\}
    \\&+ (1-\mu|\psi_\tc{c}|^2) \partial^\mu T^\textrm{fluid}_{\mu\nu} -\mu T^\textrm{fluid}_{\mu\nu} \partial^\mu |\psi_\tc{c}|^2 + \mu  \mathcal{L}^\textrm{fluid} \partial_\nu|\psi_\tc{c}|^2,
\end{aligned}
\label{diverce}
\end{equation}\end{widetext}
where we added and subtracted $ \mu  \mathcal{L}^\textrm{fluid} \partial_\nu|\psi_\tc{c}|^2$ to {explicitly} factor the equation of motion for the field $\psi_\tc{c}$. Using the equations of motion \eqref{eqmotion}, the first line in Eq.~(\ref{diverce}) vanishes, and we see that the divergencelessness of $T_{\mu\nu}$ is ensured provided that the perfect fluid stress-energy tensor satisfies
\begin{equation}\label{eq:Tmunurequirement}
\begin{aligned}
    (1-\mu|\psi_\tc{c}|^2) \partial^\mu T^\textrm{fluid}_{\mu\nu} &-\mu T^\textrm{fluid}_{\mu\nu} \partial^\mu |\psi_\tc{c}|^2  \\&+ \mu  \mathcal{L}^{\textrm{fluid}} \partial_\nu|\psi_\tc{c}|^2 = 0,
\end{aligned}\end{equation}
which turns into a differential equation for $u_\mu$, $\rho$ and $P$.

We are interested in using the field $\psi_\tc{c}$ to source a time-independent potential for the field $\phi_\tc{d}$. This can be obtained if $\psi_\tc{c}$ is of the form
\begin{equation}\label{eq:ansatzpsic}
    \psi_\tc{c}(\mf x) = e^{\ii \omega_\tc{c} t}\Psi_\tc{c}(\bm x).
\end{equation}
Below we will analyze a particular case which allow this ansatz for $\psi_\tc{c}$ to be a solution to the equations of motion.

\subsection*{Time Independent Solutions}
\label{time independent solutions}

We will now analyze the case where the field $\phi_\tc{d}$ is such that $\phi_\tc{d}^2(\mf x) = g(\bm x)$ is time independent in a given inertial frame. In the semiclassical context $\phi_\tc{d}^2(\mf x)$ would be replaced by the renormalized expected value of $\hat{\phi}_\tc{d}^2$,  $\langle \hat{\phi}_\tc{d}^2(\mf x)\rangle_{\text{ren}}$, which is time independent whenever $\hat{\phi}_\tc{d}$ is in an eigenstate of its Hamiltonian (e.g., if $\alpha = 0$ and $\hat{\phi}_\tc{d}$ were in its vacuum state, we would have $g(\bm x) = 0$).  In this case, the equation of motion for $\psi_\tc{c}(\mf x) = e^{\ii \omega_\tc{c} t}\Psi_\tc{c}(\bm x)$ reads
\begin{equation}\begin{aligned}
    \Big(\omega_\tc{c}^2 - m_\tc{c}^2& + \nabla^2 - f(\bm x)\Big)\Psi_\tc{c}(\bm x) = 0,
    \end{aligned}
\end{equation}
where we defined
\begin{equation}\label{eq:f}
    f(\bm x) = \mu \mathcal{L}^\textrm{fluid} + \frac{\alpha}{2}g(\bm x) + F_\tc{c}(\bm x).
\end{equation}
Notice that once $g(\bm x)$ is fixed and the potential $V_\tc{c}$ is chosen, the fluid Lagrangian completely determines $f(\bm x)$. Thus, $\Psi_\tc{c}(\bm x)$ is an eigenfunction of the operator \mbox{$-\nabla^2 + f(\bm x)$}. Recalling that we are interested in localized solutions, we should
look for eigenfunctions with negative eigenvalues:
\begin{equation}
    (-\nabla^2 + f(\bm x)) \Psi_\tc{c}(\bm x)  = -\lambda_\tc{c}^2 \Psi_\tc{c}(\bm x){.}
    \label{potential f}
\end{equation}
Hence, Eq. \eqref{eq:ansatzpsic} is a stationary localized 
solution to the equations of motion provided that
\begin{equation}
    \omega_\tc{c}^2 = m_\tc{c}^2 - \lambda_\tc{c}^2,
    \label{frequency}
\end{equation}
with $m_\tc{c}^2 > \lambda_\tc{c}^2$. 

Due to the fact that $|\psi_\tc{c}(\mf x)|^2 = |\Psi_\tc{c}(\bm x)|^2$ is time independent in this case, it is natural to impose that the fluid described by $T^\textrm{fluid}_{\mu\nu}$ undergoes motion in the $\partial_t$ direction, and that both $\rho$ and $P$ are time-independent. We then find that the $0-$th component of Eq.~(\ref{eq:Tmunurequirement}) is trivial, whilst 
\begin{equation}
    \partial^\mu T^\textrm{fluid}_{\mu i } = \partial_i  P
\end{equation}
and Eq. \eqref{eq:Tmunurequirement} becomes a differential equation for $P$, 
\begin{equation}
\begin{aligned}
    &(1-\mu |\Psi_\tc{c}|^2)\partial_i P - \mu  \partial_i|\Psi_\tc{c}|^2 P \\
    &+(f(\bm x) -\tfrac{\alpha}{2}g(\bm x) -F_\tc{c}(\bm x))\partial_i |\Psi_\tc{c}|^2 = 0
\end{aligned}
\label{eq for P}
\end{equation}
where $f$, $g$, and $|\Psi_\tc{c}|^2$ are given.

Finding a solution $P$ for the above equation also gives us the proper energy density of the fluid ($\rho$) through the equation $\mu \mathcal{L}^\textrm{fluid} + \frac{\alpha}{2}g(\bm x) +F_\tc{c}\left(|\psi_\tc{c}|^2\right)= f(\bm x)$. In particular, the proper energy density of the fluid depends on the choice of the on-shell Lagrangian $\mathcal{L}^\textrm{fluid}$ (i.e., the choice of the non-minimal coupling between the fluid and $\psi_\tc{c}$).

We consider two on-shell Lagrangians given by \mbox{$\mathcal{L}^\textrm{fluid}=-\rho+3\eta P$} so that \mbox{$\eta=0$} for the choice \mbox{$\mathcal{L}^\textrm{fluid}=-\rho$} and $\eta=1$ for \mbox{$\mathcal{L}^\textrm{fluid}=T^{\textrm{fluid}}=-\rho+3P$}~\footnote{We will not consider Lagrangians that are independent of $\rho$ (such as $\mathcal{L}^\textrm{fluid} = P$), so that the energy density can be determined from $\mathcal{L}^\textrm{fluid}$.}.  We can then find the equation of state of the fluid, relating $\rho$ and $P$, in terms of the functions $g(\bm x)$, $F_\tc{c}(\bm x)$ and $f(\bm x)$: 
\begin{equation}
    \rho = 3 \eta P + \frac{\alpha}{2 \mu}g(\bm x) + \frac{F_{\tc{c}}(\bm x)}{\mu}  - \frac{f(\bm x)}{\mu}.
    \label{eq for rho}
\end{equation}

For a given time independent configuration of the field $\phi_\tc{d}$, this solution is stable, and satisfies $\partial^\mu T_{\mu\nu} = 0$ in the whole spacetime. In particular, notice that the function $f(\bm x)$ is independent of the choice of $g(\bm x)$, so that different $g$'s yield different pressures $P$ and proper energy densities $\rho$ for the fluid. Also notice that whether $T^{\textrm{fluid}}_{\mu\nu}$ satisfies energy conditions or not will explicitly depend on the choices of $f(\bm x)$, $g(\bm x)$ and $\Psi_\tc{c}(\bm x)$. Ideally, to ensure localization of the whole system (in the sense that $T_{\mu\nu}$ goes to zero at spatial infinity), one would require that both $\rho$ and $P$ go to zero at infinity, so that the integration constant arising in Eq. \eqref{eq for P} is not arbitrary.

\section{An explicit example of a localized quantum field}
\label{sec:example}

In this section we will present a concrete example of a realization of the time independent model constructed in Section \ref{time independent solutions}. As discussed above, the first step is to pick the state of the classical field which will source the external potential for the quantum field $\phi_\tc{d}(\mf x)$. The explicit example we will construct will be spherically symmetric, so we use spherical coordinates $(t,r,\theta,\phi)$, and choose the state for the field $\psi_\tc{c}(\mf x)$ to be
\begin{equation}
    \psi_\tc{c}(\mf x) = \frac{1}{\ell}e^{-\ii \omega_\tc{c} t}\sech(\frac{r}{\ell}),
    \label{psic}
\end{equation}
so that the effective potential generated by $\psi_\tc{c}(\mf x)$ on $\phi_\tc{d}$ is given by
\begin{equation}
    \alpha|\psi_\tc{c}|^2 = \frac{\alpha}{\ell^2} \sech^2\left(\frac{r}{\ell}\right).
\end{equation}
This implies that the equation of motion for the field $\phi_\tc{d}$ becomes
\begin{equation}
    \left(\Box-m_\tc{d}^2 - \frac{\alpha}{\ell^2}\sech^2\left(\tfrac{r}{\ell}\right)\right)\phi_\tc{d}  = 0.
\end{equation}
We also find that
\begin{equation}
    \nabla^2 \Psi_\tc{c}(\bm x)  = \left(\frac{1}{\ell^2} -\frac{2}{\ell^2} \sech^2\left(\tfrac{r}{\ell}\right) - \frac{2}{\ell^2} \frac{\tanh(\tfrac{r}{\ell})}{\tfrac{r}{\ell}}\right)\Psi_\tc{c}(\bm x),
    \label{eqpsic}
\end{equation}
so that it has the shape of the eigenvalue equation~\eqref{potential f} with
\begin{align}
    f(\bm x) &= f(r) = -\frac{2}{\ell^2} \sech^2(\tfrac{r}{\ell}) - \frac{2}{\ell^2} \frac{\tanh(\tfrac{r}{\ell})}{r/\ell},
\end{align}
and the corresponding eigenvalue $\lambda_{\tc{c}} = 1/\ell$, so that \mbox{$(\nabla^2 + f(\bm x)) \Psi_\tc{c}= \lambda_{\tc{c}}\Psi_\tc{c}$}. We also find the frequency $\omega_\tc{c}$ from Eq. \eqref{frequency}
\begin{equation}
    \omega_{\tc{c}} = \sqrt{m_\tc{c}^2 - \frac{1}{\ell^2}}.
\end{equation}
For convenience, we assume that $\langle:\!\hat{\phi}_\tc{d}^2(\mf x)\!:\rangle = g(\mf x) = 0$ here so that we can pick $\mathcal{L}^\textrm{fluid}$ and the potential $V_\tc{c}(|\psi_\tc{c}|^2)$ as
\begin{equation}
    V_\tc{c}(|\psi_\tc{c}|^2) = -(|\psi_\tc{c}|^2)^2, \quad\quad \mathcal{L}^{\textrm{fluid}} =-\frac{2}{\mu\ell^2} \frac{\tanh(\tfrac{r}{\ell})}{r/\ell},
\end{equation}
so that $F_\tc{c}(\bm x)$ is given by
\begin{equation}
    F_\tc{c}(\bm x) = -2|\psi_\tc{c}(\mf x)|^2 = -\frac{2}{\ell^2}\sech^2(\tfrac{r}{\ell}),
\end{equation}
and Eq.~\eqref{eq:f} is satisfied. Notice that both $\mathcal{L}^{\text{fluid}}$ and $V_\tc{c}(\bm x)$ are both smooth bounded functions in space. Equation \eqref{potential f} then holds provided that $\psi_\tc{c}(\mf x)$ is given by Eq. \eqref{psic}. 

Now one would need to solve Eq. \eqref{eq for P} for the pressure $P$ and Eq. \eqref{eq for rho} for the energy density $\rho$. Due to spherical symmetry and time invariance of the system, we have $P = P(r)$ and $\rho = \rho(r)$. The differential equation for $P$ becomes
\begin{equation}
    P'(r)  + \frac{2 \mu \tanh(\tfrac{r}{\ell})P(r)}{\ell(-\mu + \ell^2\cosh^2(\tfrac{r}{\ell}))}  = \frac{4 \tanh^2(\tfrac{r}{\ell})}{r\ell^2( - \mu+\ell^2 \cosh^2(\tfrac{r}{\ell}))}.
\end{equation}
The solution for $P(r)$ can be written as
\begin{align}\label{eq:P}
    P(r)  &= \frac{1}{\ell^2\left(\ell^2 -\mu \sech^2(\tfrac{r}{\ell})\right)}\int_r^\infty dr' G(r'),
\end{align}
where
\begin{align}
    G(r) \label{eq:Gofr}
    &=\frac{4 \sech^2(\tfrac{r}{\ell})\tanh^2(\tfrac{r}{\ell})}{r},
\end{align}
and we picked the integration constant such that $\lim_{r\to \infty} P(r) = 0$. Unfortunately, no known closed expression for the integral of $G(r)$ is known. The energy density can be found by using Eq.~\eqref{eq for rho}, which yields
\begin{equation}\label{eq:rho}
    \rho(r) = 3\eta  P(r) + \frac{2}{\mu\ell^2}\frac{\tanh(\tfrac{r}{\ell})}{r/\ell}. 
\end{equation}

In order to have a concrete model, we will also pick a value for the constant $\mu$. However, not all values for $\mu$ will yield a physical model. Notice that for $\mu  >\ell^2$, the solution for $P(r)$ is divergent at $r = \ell \,\text{arcsech}(\ell^2/|\mu|)$. Provided that $\mu <\ell^2$, the solution for $P(r)$ is smooth, positive and decreasing. It is also important to ensure that the energy density of the fluid is positive. First, notice that for large $r$, $\rho(r)$ behaves as $\frac{2}{\mu \ell^2 r}$, so that we must have $\mu>0$ to ensure $\rho(r)>0$. Given that $P(r)$ is positive and smooth for $\mu<\ell^2$, $\mu \in (0,\ell^2)$ ensures that $\rho(r)>0$ for all $r$.

\begin{figure}[htb]
    \centering
    \includegraphics[width=8.6cm]{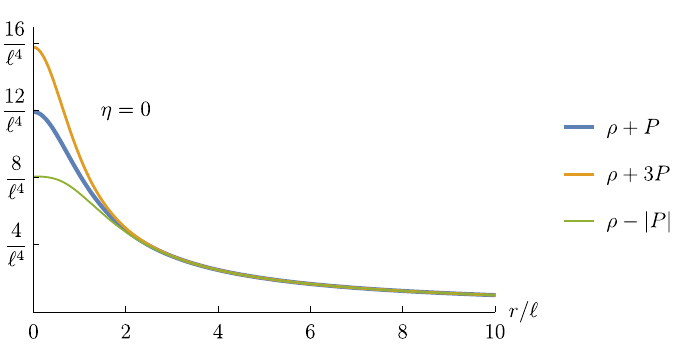}
    \caption{$\rho+P$, $\rho+3P$, $\rho-|P|$ as a function of $r/\ell$ for $\eta=0$ and $\mu=\ell^2/5$.}
    \label{fig1}
\end{figure}

\begin{figure}[htb]
    \centering
    \includegraphics[width=8.6cm]{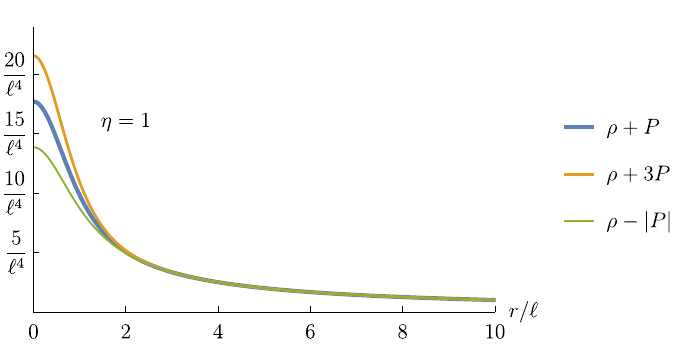}
    \caption{$\rho+P$, $\rho+3P$, $\rho-|P|$ as a function of $r/\ell$ for $\eta=1$ and $\mu=\ell^2/5$.}
    \label{fig2}
\end{figure}

One can go a step further and demand that the null, weak, strong and dominant energy conditions are satisfied by the fluid, imposing that $\rho+P>0$, $\rho+3P>0$ and $\rho-|P|>0$. Given that both $\rho$ and $P$ are positive, the only condition that adds extra constraints is $\rho-|P|>0$. We find that $\rho-|P|>0$ if $2/\mu > 3 (\eta-1) P(0)$, and $P(0)$ can be computed in closed form~\cite{mathematica}:
\begin{align}
    P(0) &= \frac{4 \left(4 \log(A) - 40 \zeta'(-3) - \frac{1}{3} - \frac{4}{45}\log(2)\right)}{\ell^2(\ell^2 - \mu)}\\
    & \equiv \frac{g_0}{\ell^2(\ell^2  - \mu)}
\end{align}
where $\zeta$ denotes Riemann's Zeta function and $A$ is the Glaisher constant, yielding $g_0 \approx 1.53971$. We then find that $\rho(r) - |P(r)|>0$ if either $\eta>\frac{1}{3}$, or if $\mu < \frac{\ell^2}{1+(1-3\eta) g_0/2}$. 
For instance, if $\eta=0$ we have that $\rho(r) - |P(r)|$ will only be positive if $0<\mu\lesssim 0.565017 \ell^2$. In Figs.~\ref{fig1} and~\ref{fig2} we plot $\rho+P$, $\rho+3P$ and $\rho-|P|$ for values of the constant $\mu$ that respect the energy conditions.
\begin{figure}[htb]
    \centering
    \includegraphics[width=8.6cm]{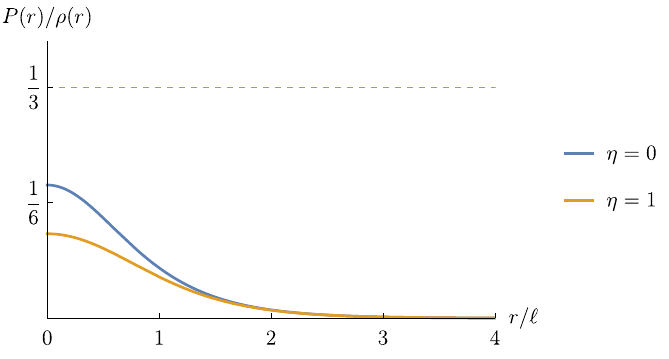}
    \caption{$w:=P/\rho$ for $\eta=0$ and $\eta=1$ with the choice $\mu = \ell^2/5$.}
    \label{figw}
\end{figure}
Fig.~\ref{figw} shows that $0<w:=P/\rho<1/3$ on both cases $\eta=0$ and $\eta=1$. This is specially important for the case $\eta=1$, where the fluid is constituted by particle with fixed mass and structure. In any case, the fluid is composed of non-exotic matter content. 
Finally, as expected, both quantities are localized around the origin within a lengthscale characterized by the parameter $\ell$.

With these classical solutions for the fields $\psi_\tc{c}(\mf x)$, and for the fluid parameters $\rho(r)$ and $P(r)$, the field $\phi_{\tc{d}}$ experiences the effective potential 
\begin{equation}
    V(r) = \frac{\alpha}{\ell^2} \sech^2\left(\frac{r}{\ell}\right).
\end{equation}
From here on, we pick $\alpha = -6$. The equation of motion for the field $\phi_\tc{d}$ can then be solved via separation of variables. The operator
\begin{equation}
    L = - \nabla^2 - \frac{6}{\ell^2}\sech^2(\tfrac{r}{\ell}) 
\end{equation}
possesses one eigenfunction with negative eigenvalue and a continuous spectrum in $[0,\infty)$. The Klein-Gordon normalized eigenfunction of $L$ and its respective eigenvalue are
\begin{align}
    \Phi_1(r) &= \sqrt{\frac{3}{8\pi \ell\omega_\tc{d}}}\frac{\tanh(\tfrac{r}{\ell})}{r \cosh(\tfrac{r}{\ell})}, &&& \mu_1 &= -\frac{1}{\ell^2}.\label{eq:Phi1}
\end{align}
We label the generalized eigenfunctions in the continuous spectrum of $L$ by $v_{klm}(\bm x)$, where $k^2/\ell^2$ is the corresponding eigenvalue, and $l,m$ are the usual angular momentum labels.

The corresponding quantum field $\hat{\phi}_\tc{d}$ can then be written as
\begin{align}
    \hat{\phi}_\tc{d}(\mf x) =& \left(e^{- \ii \omega_{1} t}\Phi_{1}(r) \hat{a}_{1} + e^{\ii \omega_{1} t}\Phi^*_{1}(r) \hat{a}^\dagger_{1}\right)\\
    &\!\!\!\!\!\!\!\!\!\!\!\!\!\!\!\!\!+ \sum_{l,m}\int\dd k \left(e^{- \ii w_{k} t}v_{klm}(\bm x) \hat{b}_{klm} +e^{\ii w_{k} t}v^*_{klm}(\bm x) \hat{b}^\dagger_{klm} \right),\nonumber
\end{align}
where $\omega_1 = \sqrt{m_\tc{d}^2 - \tfrac{1}{\ell^2}}$ and $w_{k} = \sqrt{m_{\tc{d}}^2 + \frac{k^2}{\ell^2}}$. Notice that the modes $v_{klm}(\bm x)$ are not localized, and do not belong to $L^2(\mathbb{R}^3)$, as they are associated with scattering states.

\section{The stress-energy tensor of a UDW detector}
\label{sec:tmunuexample}

In this section we describe the stress-energy tensor of a UDW detector that matches the description of Section~\ref{sec:example}. The energy-tensor for this configuration can be written as Eq.~\eqref{eq:fullTmunu}, with the replacement $\phi_\tc{d}(\mf x)\mapsto \hat{\phi}_\tc{d}(\mf x)$, giving rise to an operator-valued energy momentum tensor $\hat{T}_{\mu\nu}$. One can then obtain a classical stress-energy tensor by taking the expected value of the renormalized energy tensor
\begin{equation}
    \langle:\!\hat{T}_{\mu\nu}\!:\rangle_{\hat{\rho}_\tc{d}} = \langle\hat{T}_{\mu\nu}\rangle_{\hat{\rho}_\tc{d}} - \bra{0_\tc{d}}\hat{T}_{\mu\nu}\ket{0_\tc{d}},
\end{equation}
where $\hat{\rho}_\tc{d}$ denotes the state of the detector field $\hat{\phi}_\tc{d}$ and we choose to use as a reference state the vacuum\footnote{In order to obtain a finite value for the stress-energy tensor, one requires to choose a renormalization scheme. We choose the reference state $\ket{0_\tc{d}}$ for convenience, but another natural choice would be to consider a subtraction using the Minkowski vacuum.} of the field $\hat{\phi}_\tc{d}(\mf x)$. For convenience, we define
\begin{align}
    \hat{T}_{\mu\nu}^{\phi_\tc{d}} &=  \partial_\mu \hat{\phi}_{\tc{d}}\partial_\nu \hat{\phi}_{\tc{d}} -\frac{1}{2}g_{\mu\nu}\!\left(\partial_\alpha \hat{\phi}_{\tc{d}}\partial^\alpha \hat{\phi}_{\tc{d}} + m_{\tc{d}}^2 \hat{\phi}_\tc{d}^2\right)\!, \\
    \hat{T}_{\mu\nu}^{\phi_\tc{d}\psi_\tc{c}} &= -\frac{1}{2}g_{\mu\nu}\alpha |\psi_\tc{c}|^2 \hat{\phi}_\tc{d}^2,\\
    T_{\mu\nu}^{\psi_\tc{c}} &= 2\Re(\partial_\mu \psi_{\tc{c}}^*\partial_\nu \psi_{\tc{c}})\\&\:\:\:\:\:-g_{\mu\nu}\left(\partial_\alpha \psi_{\tc{c}}^*\partial^\alpha \psi_{\tc{c}} + m_{\tc{c}}^2 |\psi_\tc{c}|^2  +  V_\tc{c}(|\psi_\tc{c}|^2)\right),\nonumber\\
    T_{\mu\nu}^{\psi_\tc{c}\text{fluid}} &=  -\mu |\psi_\tc{c}|^2T^\textrm{fluid}_{\mu\nu},
\end{align}
so that the stress-energy tensor of the detector can be expressed as the sum
\begin{equation}
    \hat{T}_{\mu\nu} = \hat{T}_{\mu\nu}^{\phi_\tc{d}} + \hat{T}_{\mu\nu}^{\phi_\tc{d}\psi_\tc{c}} + T_{\mu\nu}^{\psi_\tc{c}} + T_{\mu\nu}^{\psi_\tc{c}\text{fluid}} + T^\textrm{fluid}_{\mu\nu}.
\end{equation}
Notice that only $\hat{T}_{\mu\nu}^{\phi_\tc{d}}$ and $\hat{T}_{\mu\nu}^{\phi_\tc{d}\psi_\tc{c}}$ are operator valued. However, all terms that contain dependence on the fluid indirectly depend on the state of the field $\hat{\phi}_\tc{d}$ through the function $g(\bm x)$.

As an explicit example, we consider a Harmonic oscillator UDW detector interacting with a free massless quantum field according to the interaction Hamiltonian of Eq.~\eqref{eq:HIfield}. We consider the detector to have the spacetime smearing function
\begin{equation}
    \Lambda(\mf x) = e^{-\frac{t^2}{2T^2}} \sqrt{\frac{3}{8\pi \ell\omega_{\tc{d}}}}\frac{\tanh(\tfrac{r}{\ell})}{r \cosh(\tfrac{r}{\ell})}.
\end{equation}
This spacetime smearing function corresponds to a detector modelled by the bound mode $\Phi_1(r)$ in Eq.~\eqref{eq:Phi1} and a switching function $\zeta(\mf x) = e^{-\frac{t^2}{2 T^2}}$ that controls the time profile of the interaction with the parameter $T$. This detector is well modelled (to leading order) by the field $\hat{\phi}_\tc{d}$, interacting with the field $\hat{\phi}$, as discussed in the previous section.

\begin{figure}[htb]
    \centering
    \includegraphics[width=8.6cm]{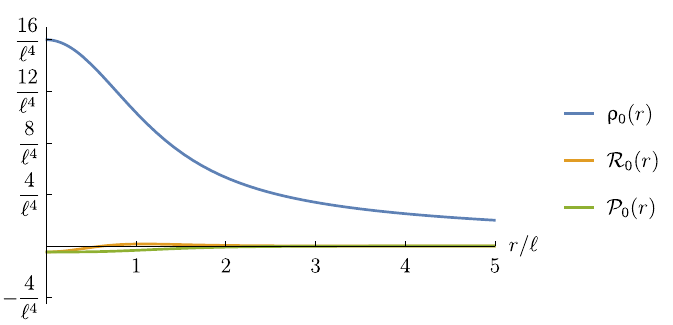}
    \caption{$\uprho_0(r), \mathcal{R}_0(r),$ and $\mathcal{P}_0(r)$ for $\eta=0$, $\mu=\ell^2/5$, $m_\tc{c} = \frac{2}{\ell}$.}
    \label{Tmunu0}
\end{figure}

The detector is modelled by the combination of the field $\psi_\tc{c}$, the fluid, and the field $\hat{\phi}_\tc{d}$, which models its internal dynamics. When the detector is in its ground state, the field $\hat{\phi}_\tc{d}$ is then in its vacuum state,  $\ket{0_\tc{d}}$. We assume that the free field $\hat{\phi}$ is in a state denoted by $\hat{\rho}_\phi$. The stress-energy tensor of the detector is then a combination of the stress-energy tensor of the quantum field $\hat{\phi}_\tc{d}$, the classical field $\psi_\tc{c}$, and the perfect fluid. In the vacuum state $\ket{0_\tc{d}}$, we have $\langle: \!\hat{T}_{\mu\nu}^{\phi_\tc{d}}\!:\rangle = \langle:\!\hat{T}_{\mu\nu}^{\phi_\tc{d}\psi_\tc{c}}\!:\rangle = 0$, so that the stress-energy tensor of this system in vacuum state, $T_{\mu\nu}^{\ket{0_\tc{d}}} = \bra{0_\tc{d}} \!:\hat{T}_{\mu\nu}:\!\ket{0_\tc{d}}$, can then be written as
\begin{equation}
    T_{\mu\nu}^{\ket{0_\tc{d}}} = T_{\mu\nu}^{\psi_\tc{c}} + T_{\mu\nu}^{\psi_\tc{c}\text{fluid}} + T^\textrm{fluid}_{\mu\nu},
\end{equation}
with $\rho$ given by Eq.~\eqref{eq:rho}, $P$ given by Eq.~\eqref{eq:P} and $\psi_\tc{c}(\mf x)$ given by Eq.~\eqref{psic}. This results in a $T_{\mu\nu}$ of the form
\begin{equation}
    T_{\mu\nu}^{\ket{0_\tc{d}}} = \uprho_0(r)u_\mu u_\nu + \mathcal{R}_0(r)r_\mu r_\nu + \mathcal{P}_0(r) \Omega_{\mu\nu},\label{eq:Tmunu0}
\end{equation}
where $\mathsf{u} = \mathsf{e}_t$, $\mathsf{r} = \mathsf{e}_r$, $\Omega = \mathsf{e}_\theta\otimes \mathsf{e}_\theta + \mathsf{e}_\phi\otimes \mathsf{e}_\phi$ in the normalized spherical frame $\mf{e}_t = \partial_t$, $\mf{e}_r = \partial_r$, $\mf{e}_\theta = \frac{1}{r}\partial_\theta$, $\mf{e}_\phi = \frac{1}{r \sin\theta}\partial_\phi$. The energy tensor is then diagonal in spherical coordinates, and $\uprho_0(r) = T^{\mu\nu}_0u_\mu u_\nu$ corresponds to the energy density in this frame, $\mathcal{R}_0(r)  = T^{\mu\nu}r_\mu r_\nu$ is the radial pressure, and $\mathcal{P}_0(r) = T^{\mu\nu} \Omega_{\mu\nu}$ is the pressure in the angular directions. Their explicit expressions are given below
\begin{align}
    \uprho_0(r) &= \frac{2\sech^2(\tfrac{r}{\ell})}{\ell^2}m_\tc{c}^2  + \left(1 - \frac{\mu \sech^2(\tfrac{r}{\ell})}{\ell^2}\right)\rho(r),\\
    \mathcal{R}_0(r) &= - \frac{2\sech^4(\tfrac{r}{\ell})}{\ell^4}+ \left(1 - \frac{\mu \sech^2(\tfrac{r}{\ell})}{\ell^2}\right)P(r),\nonumber\\
    \mathcal{P}_0(r) & = -\frac{2\sech^2(\tfrac{r}{\ell})}{\ell^4} + \left(1 - \frac{\mu \sech^2(\tfrac{r}{\ell})}{\ell^2}\right)P(r) \nonumber.
\end{align}
The plots of $\uprho_0(r), \mathcal{R}_0(r),$ and $\mathcal{P}_0(r)$ can be found in Fig.~\ref{Tmunu0}. We see that the radial and angular pressures are negative assume negative values, but it is simple to check that all energy conditions are verified. Notice that these results are independent of any specific property of the field $\hat{\phi}_\tc{d}$, as they correspond only to the system responsible for generating the trapping potential (which depends on $m_\tc{c}$, $\ell$, and the parameters of the fluid).

\begin{figure}[htb]
    \centering
    \includegraphics[width=8.6cm]{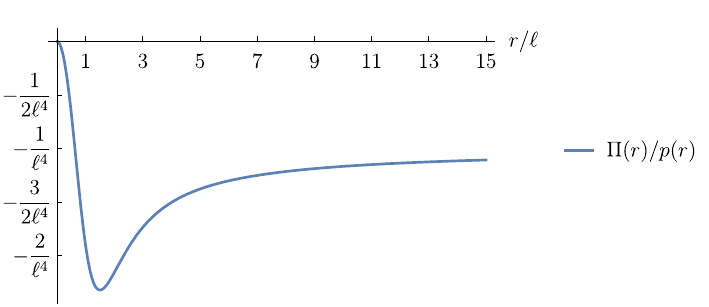}
    \caption{The pressure deviator $\Pi(r)/p(r)$ for $\eta = 0=$, $\mu = \ell^2/5$, $m_\tc{c} = \frac{2}{\ell}$.}
    \label{deviator}
\end{figure}

Given the pressures $\mathcal{R}_0(r)$ and $\mathcal{P}_0(r)$ we can calculate the pressure deviator $\Pi(r)$~\cite{landau}. This quantity is defined as the traceless part of the the spatial components of the energy momentum tensor. It measures the difference  from the matter content described by $\uprho_0(r)$,  $\mathcal{R}_0(r)$ and $\mathcal{P}_0(r)$ and a perfect fluid modelled by a gas of particles. It can be calculated through the Landau decomposition~\cite{landau}
\begin{equation}
    T_{\mu\nu} = \rho u_\mu u_\nu + (p(r) +\Pi(r))r_\mu r_\nu + (p(r) - \tfrac{1}{2}\Pi(r))\Omega_{\mu\nu},
\end{equation}
where
\begin{align}
    \Pi(r) &= \frac{2}{3}\left(\mathcal{R}_0(r) - \mathcal{P}_0(r)\right) = \frac{4}{3}\frac{\sech^2(\tfrac{r}{\ell})\tanh^2(\tfrac{r}{\ell})}{\ell^4},\nonumber\\
    p(r) &= \frac{\mathcal{R}_0(r) + 2 \mathcal{P}_0(r)}{3}.
\end{align} 
In Fig.~\ref{deviator}, we plot $\Pi(r)/p(r)$ as a function of $r/l$. We observe that the radial pressure approaches zero as $r \to \infty$, indicating that $T_{\mu\nu}$ does not represent a perfect fluid, even at infinity.

\begin{figure}[htb]
    \centering
    \includegraphics[width=8.6cm]{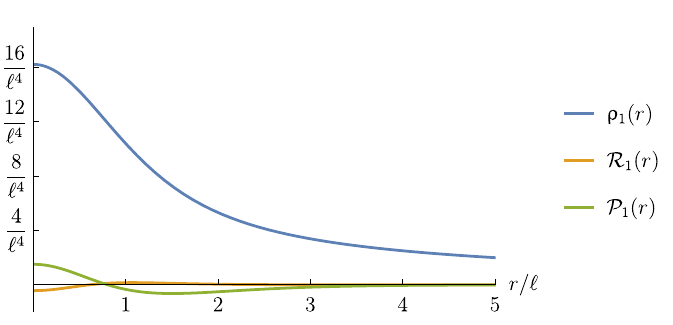}
    \caption{$\uprho_1(r), \mathcal{R}_1(r),$ and $\mathcal{P}_1(r)$ for $\eta=0$, $\mu=\ell^2/5$, $m_\tc{c} = \frac{2}{\ell}$, and $m_\tc{d} = \frac{5}{\ell}$.}
    \label{Tmunu1}
\end{figure}

The detector's first excited state is \mbox{$\ket{1_\tc{d}} = \hat{a}_1^\dagger \ket{0_\tc{d}}$}. To compute the stress-energy tensor when the detector is in this configuration, one could perform the same procedure as that of Section~\ref{sec:example}, using $g(\bm x)$ as the renormalized expected value of $\hat{\phi}_\tc{d}^2$, so that the fluid absorbs the dependence on $g(\bm x)$. We have
\begin{align}
    g(\bm x) &= \langle 1_\tc{d} | \!:\!\hat{\phi}_\tc{d}^2(\mf x)\!:\!| 1_\tc{d} \rangle =\langle 1_\tc{d} | \hat{\phi}_\tc{d}^2(\mf x)| 1_\tc{d} \rangle - \langle 0_\tc{d} | \hat{\phi}_\tc{d}^2(\mf x)| 0_\tc{d} \rangle\nonumber \\
    &= \frac{6\csch^4(\tfrac{2r}{\ell})\sinh^6(\tfrac{r}{l})}{\pi r^2 \omega_\tc{d} \ell}.
\end{align}
The energy density and pressure of the perfect fluid are then changed to $\rho_1(r)$ and $P_1(r)$, given explicitly by
\begin{equation}
    P_1(r) = \frac{1}{\ell^2 - \mu \sech^2(\tfrac{r}{\ell})}\int_r^\infty G_1(r),
\end{equation}
where $G_1(r) = G(r) + \Delta G(r)$ with $G(r)$ defined in Eq.~\eqref{eq:Gofr}, and 
\begin{align}
    \Delta G(r) &= -\frac{9 \tanh^3(\tfrac{r}{\ell})\sech^4(\tfrac{r}{\ell})}{4 \pi r^2 \ell^2 \omega_\tc{d}}.
\end{align}
The energy density will then be given by
\begin{equation}
    \rho_1(r) = 3 \eta P_1(r) - \mathcal{L}^{\text{fluid}},
\end{equation}
where
\begin{align}
    \mathcal{L}^\text{fluid} =-\frac{2}{\mu\ell^2} \frac{\tanh(\tfrac{r}{\ell})}{r/\ell} - \frac{\alpha}{2\mu}g(\bm x).
\end{align}
Computing the expected value of the renormalized stress-energy densities $\langle: \!\hat{T}_{\mu\nu}^{\phi_\tc{d}}\!:\rangle$ and $\langle:\!\hat{T}_{\mu\nu}^{\phi_\tc{d}\psi_\tc{c}}\!:\rangle$, one obtains a stress energy tensor of the form
\begin{equation}
    T^{\ket{1_\tc{d}}}_{\mu\nu} = \uprho_1(r)u_\mu u_\nu + \mathcal{R}_1(r)r_\mu r_\nu + \mathcal{P}_1(r) \Omega_{\mu\nu},
\end{equation}
where $\uprho_1(r)$, $\mathcal{R}_1(r)$, and $\mathcal{P}_1(r)$ play the same role as $\uprho_0(r)$, $\mathcal{R}_0(r)$, and $\mathcal{P}_0(r)$ in Eq.~\eqref{Tmunu0}. However, their expressions are cumbersome and do not provide any important insight. The plots of these quantities (when the detector is in its excited state) is displayed in Fig.~\ref{Tmunu1}.

\begin{figure}[h!]
    \centering
    \includegraphics[width=8.6cm]{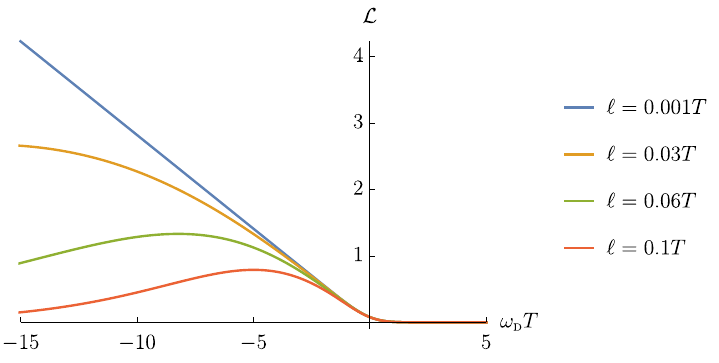}
    \caption{The leading order excitation probability of the detector as a function of $\omega_\tc{d} T$ for different values of $\ell$, which determine the detector size.}
    \label{excitationProb}
\end{figure}

A detector that starts in its ground state evolves according to the unitary time evolution operator
\begin{equation}
    \hat{U}_I = \mathcal{T}\exp(-\ii \int \dd V \hat{\mathcal{H}}_I(\mf x)).
\end{equation}
In the case where the field $\hat{\phi}_\tc{d}$ starts in the state $\ket{0_\tc{d}}$ and the field $\hat{\phi}$ starts in its vacuum state $\ket{0}$, the final state of the detector takes the form
\begin{equation}
    \hat{\rho}_\tc{d} = (1 - \lambda^2 \mathcal{L})\ket{0_\tc{d}}\!\!\bra{0_\tc{d}} + \lambda^2\mathcal{L} \ket{1_\tc{d}}\!\!\bra{1_\tc{d}} + \mathcal{O}(\lambda^4),
\end{equation}
where
\begin{equation}
    \mathcal{L} = \int \dd V \dd V' \Lambda(\mf x) \Lambda(\mf x') e^{- \ii \Omega (t-t')}\langle 0|\hat{\phi}(\mf x) \hat{\phi}(\mf x')|0\rangle
\end{equation}
is the leading order excitation probability of the detector, shown in Fig.~\ref{excitationProb}. In the plot we see that in the pointlike limit, we approximate the excitation probability of a pointlike UDW detector, given by
\begin{equation}
    \mathcal{L}(\Omega) \approx \frac{ |\Omega| T}{2\sqrt{\pi}}\Theta(-\Omega),
\end{equation}
where $\Omega$ is the detector's energy gap $\Omega = \omega_\tc{d}$ in this case.

Given that the leading order final state of the detector is a mixture between excited and ground state, the leading order stress-energy tensor of the final state of the detector is also a mixture, given by
\begin{equation}
    T^{\mu\nu}_\tc{d} = \langle \hat{T}^{\mu\nu}\rangle_{\hat{\rho}_\tc{d}} = (1-\lambda^2\mathcal{L})T^{\mu\nu}_0 + \lambda^2 \mathcal{L}T^{\mu\nu}_1 + \mathcal{O}(\lambda^4).
\end{equation}
The specific value of the stress-energy tensor above depends explicitly on the value of the coupling strength and on the excitation probability.

\section{Conclusions}
\label{sec:conclusion}




In general relativity, a covariant description of a localized system in spacetime requires a dynamic representation of the mechanisms that localize it. This requirement arises because all forms of energy and momentum inherently couple to gravity in this framework. To address this, we introduced a model of a localized particle detector that preserves general covariance with  a covariantly conserved energy-momentum tensor.

In this model, the detector’s internal energy levels are represented by square-integrable excitations of a real scalar quantum field $\hat{\phi}_{\tc{d}}$. The localization of $\hat{\phi}_{\tc{d}}$ is achieved through a spherically symmetric smooth bounded potential produced by a classical complex field, $\psi_{\tc{c}}$. The spatial profile of $\psi_{\tc{c}}$ is stabilized by its coupling with a perfect fluid, which, with two degrees of freedom, semiclassically absorbs any backreaction from the quantum field $\hat{\phi}_{\tc{d}}$. Additionally, we demonstrate that the complete system satisfies the standard energy conditions, ensuring that the detector's stress-energy tensor consists of non-exotic matter with distinct radial and angular pressures.

The model presented here is an important step to the connection between the algebraic~\cite{FewsterVerch,fewster2} and operational~\cite{chicken} formulations of measurements in QFT in curved spacetimes, as well as to gravitational backreaction in relativistic quantum information protocols. For instance, quantum energy teleportation~\cite{teleportation,teleportation2014,nichoTeleport,teleportExperiment} has been shown to generate negative energy densities~\cite{nichoTeleport}. 
Overall, having a covariantly conserved description of probes in QFT is the first step to tackle important problems that involve both gravity and quantum field theory from an operational perspective.




\acknowledgements

The authors thank Jorma Louko for insightful discussions. T. R. P. acknowledges support from the Natural Sciences and Engineering Research Council of Canada (NSERC) via the Vanier Canada Graduate Scholarship, as well as the long-term IQOQI Turis fellowship, which allowed this project to be made concrete in Vienna. J.\ P.\ M.\ P.\ thanks the support provided in part by Conselho Nacional de Desenvolvimento Científico e Tecnológico (CNPq, Brazil) Grant No. 311443/2021-4, and Fundação de Amparo à Pesquisa do Estado de São Paulo (FAPESP) Grant No. 2022/07958-4.
D.~V.\ thanks the Institute for Quantum Optics and Quantum Information of the Austrian Academy of Science for hosting him during his sabbatical year---when this collaboration was initiated---and the 
S\~ao Paulo State Research Foundation (FAPESP) for partial financial support under grant 2023/04827-9. Research at Perimeter Institute is supported in part by the Government of Canada through the Department of Innovation, Science and Industry Canada and by the Province of Ontario through the Ministry of Colleges and Universities. Perimeter Institute and the University of Waterloo are situated on the Haldimand Tract, land that was promised to the Haudenosaunee of the Six Nations of the Grand River, and is within the territory of the Neutral, Anishinaabe, and Haudenosaunee people.

\bibliography{references.bib}

\end{document}